\documentclass[conference]{IEEEtran}
\IEEEoverridecommandlockouts
\usepackage{cite}
\usepackage{amsmath,amssymb,amsfonts}
\usepackage{algorithmic}
\usepackage{graphicx}
\usepackage{textcomp}
\usepackage{xcolor}
\usepackage{graphicx}
\usepackage{subcaption}
\usepackage{flushend}
\usepackage{url}
\usepackage{soul} 
\def\BibTeX{{\rm B\kern-.05em{\sc i\kern-.025em b}\kern-.08em
    T\kern-.1667em\lower.7ex\hbox{E}\kern-.125emX}}
\begin{document}

\title{Differentiable Proxy Learning for Adaptive Quantization Control in H.264 Video Coding}

\author{\IEEEauthorblockN{Qihan Xu and Ivan V. Baji\'{c}\thanks{This work was supported by the SFU--Huawei Canada Joint Lab and the NSERC grant RGPIN-2021-02485.}}
\IEEEauthorblockA{\textit{School of Engineering Science, Simon Fraser University}\\
Burnaby, BC, Canada 
}
}
\maketitle

\begin{abstract}
H.264 has been the most widely used video coding format for the past two decades due to its relative simplicity, efficiency, and wide availability of software and hardware implementations. 
However, optimizing codec parameters such as the quantization parameter (QP) for specific objectives (e.g., perceptual quality or machine vision tasks) is challenging due to the non-differentiable nature of standard video codecs.
While differentiable proxies have recently been used to enable gradient-based optimization around standard codecs, their fidelity to the target codec is rarely explicitly characterized.
In this paper, we propose a differentiable proxy learning method for H.264 intra codec to enable adaptive quantization control. 
Built upon a variable-rate learned compression model, the proposed proxy is made differentiable with respect to codec QP through a soft-indexing mechanism. 
It is then trained to approximate the rate-distortion behavior of H.264 under two quantization settings: global-QP, which uses one QP per image, and spatial-QP, which assigns QPs at the macroblock level.
Using the frozen trained proxy, we develop a proxy-based adaptive quantization (AQ) framework for both perceptual optimization and machine vision tasks.
Experimental results demonstrate that the proposed proxies closely approximate the rate-distortion behavior of H.264 intra codec. 
The resulting proxy-based AQ framework consistently improves rate-task trade-offs over fixed-QP H.264 baselines, achieving BD-rate reduction of up to $17.12\%$ for semantic segmentation and $15.30\%$ for MS-SSIM.


\end{abstract}


\section{Introduction}
Image and video coding has become increasingly important due to the rapid growth of visual data. H.264 \cite{wiegand2003overview} has remained widely used for video coding over the past two decades due to its relative simplicity, high coding efficiency, and broad availability in software and hardware implementations.

For standard codecs like H.264, adaptive quantization (AQ) is a practical way to improve rate-distortion and rate-task trade-offs without changing the codec itself. However, directly optimizing quantization parameter (QP) assignments is difficult: standard codecs are non-differentiable, and the relationship between QP, bitrate, and reconstruction is highly coupled, which prevents gradient-based end-to-end optimization.

One way to address this issue is to introduce differentiable proxies for the target codec during training~\cite{said_differentiable_2022}. 
Existing works explore several strategies toward this goal. 
Some methods retain the overall codec structure while replacing non-differentiable operations such as quantization or entropy coding-related components with differentiable analytical approximations, enabling gradient propagation through the codec pipeline~\cite{guleryuz2024sandwiched, huo2025deep}.

Other approaches use neural simulation modules to approximate the output behavior of a standard codec during training. Instead of implementing a full learned codec, these methods train networks to predict the decoded reconstruction and/or bitrate of the target codec, thereby enabling gradient-based optimization while avoiding repeated codec invocation~\cite{son_enhanced_2022, qiu_codec-simulation_2021, Reich_2024_CVPR}.

More recently, learned image and video compression models~\cite{minnen2018joint,balle2018variational} have been adopted as differentiable codec proxies. These methods train neural compression models to mimic the reconstruction and bitrate behavior of standard codecs, and use the resulting differentiable proxies to optimize preprocessing, postprocessing, or downstream task performance~\cite{khan_perceptual_2025,lu_preprocessing_2024, klopp2021exploit}. 

Despite this progress, most prior work uses the differentiable proxy primarily as an auxiliary module for optimizing preprocessing, rescaling, or downstream adaptation. Much less attention has been paid to the differentiable proxy itself, in particular to its fidelity to the target codec.

In this work, we build differentialable proxies for H.264 intra codec under two quantization settings: a global-QP setting, where a single QP is assigned to the entire image, and a spatial-QP setting, where QP values are assigned at the macroblock level. 
We explicitly evaluate the fidelity of the learned proxy to the target codec in terms of rate-distortion behavior and visual characteristics. We then use the frozen differentiable proxy to train AQ networks for adaptive quantization under perceptual and machine vision objectives.
The main contributions of this work are as follows:
\begin{itemize}
    \item We propose a differentiable proxy for H.264 intra codec based on a variable-rate learned compression model, and make it differentiable with respect to codec quantization parameter through a soft indexing mechanism.
    \item We develop a proxy-based AQ framework for both settings, enabling gradient-based optimization of AQ network for perceptual and machine vision objectives.
    \item We provide an empirical evaluation of the learned differentiable proxy, focusing on its fidelity to H.264 in terms of rate-distortion behavior and visual characteristics under different quantization settings.
    \item We demonstrate that the proposed proxy can support AQ for both perceptual and machine vision task scenarios.
\end{itemize}

\section{Proposed method}

In this section, we first introduce a differentiable proxy learning method for H.264 intra codec, including the soft-indexing mechanism and the corresponding training strategy under global-QP and spatial-QP settings. We then present a proxy-based AQ framework that uses the frozen proxy to train AQ networks for machine vision and perceptual objectives.

\subsection{Differentiable proxy learning for H.264 intra codec} 
Let $\mathcal{C}$ denote the standard H.264 intra codec and
$\mathcal{P}_{\boldsymbol{\theta}}$ denote the differentiable learned proxy parameterized by $\boldsymbol{\theta}$.
Let $\mathbf{X}$ denote the input image and $\mathbf{Q}$ denote the QP map. In the global-QP setting, $\mathbf{Q}\in\mathbb{R}^{1\times1}$ reduces to a scalar $Q$ shared by the whole image; in the spatial-QP setting, $\mathbf{Q}\in\mathbb{R}^{H_b\times W_b}$ assigns QP values at the macroblock level. For an image of size $H\times W$, we partition it into $b\times b$ macroblocks with $b=16$ in this work, from which we have $H_b=H/b$ and $W_b=W/b$. The standard codec produces the reconstructed image and bitrate

\begin{equation}
(\hat{\mathbf{X}}_c, R_c) = \mathcal{C}(\mathbf{X}, \mathbf{Q}),
\end{equation}
while the differentiable proxy predicts

\begin{equation}
(\hat{\mathbf{X}}_p, R_p) = \mathcal{P}_{\boldsymbol{\theta}}(\mathbf{X}, \mathbf{Q}),
\end{equation}
The proxy is trained to approximate both the reconstruction and rate behavior of the standard codec:
\begin{equation}
\boldsymbol{\theta}^*
=
\arg\min_{\boldsymbol{\theta}}
\mathbb{E}
\left[
\mathcal{L}
\big(
\mathcal{P}_{\boldsymbol{\theta}}(\mathbf{X},\mathbf{Q}),
\mathcal{C}(\mathbf{X},\mathbf{Q})
\big)
\right],
\end{equation}
where $\mathcal{L}(\cdot)$ denotes the proxy training objective used to match the behavior of the standard codec.

Specifically, our proxy is built upon DCVC-FM\cite{li2024neural} intra codec, a variable-rate neural compression model that uses learnable quantization scaling. In DCVC-FM, different rate-distortion (RD) operating points are selected through discrete quantization indices associated with learned scaling factors. Although this design enables variable-rate compression, the discrete index selection is non-differentiable with respect to the quantization parameter. 

To adapt DCVC-FM for H.264 proxy learning, we introduce several key modifications. First, since the proxy control parameter is inherited from the discrete quality index in DCVC-FM, it is not numerically identical to the codec QP. We therefore establish a mapping from the codec QP variable $\mathbf{Q}$ to the proxy control variable $\mathbf{q}$ by matching their rate-distortion behavior. Each codec QP value is mapped to the nearest proxy index by minimizing the Euclidean distance in the BPP--PSNR space, based on average statistics measured on the Kodak dataset~\cite{kodakdataset} using the pretrained DCVC-FM intra codec. For the spatial-QP setting, the mapping is applied independently to each macroblock location. Second, we make the quantization mechanism differentiable through soft indexing to enable gradient-based optimization. Finally, we design proxy training strategies to match the RD behavior of H.264 under both global-QP and spatial-QP settings.

\subsubsection{Differentiable quantization scaling via soft indexing}

In DCVC-FM, variable-rate compression is achieved by modulating the encoder and decoder features using quantization-dependent scaling factors before entropy coding and reconstruction. To enable gradient-based optimization, we introduce a soft indexing mechanism inspired by the soft-assignment relaxation in soft-to-hard vector quantization~\cite{agustsson2017soft}. We apply the same principle to the selection of learnable quantization scaling factors, making the scaling differentiable with respect to the proxy control map $\mathbf{q}$.

Let $q_{u,v}$ denote the proxy control parameter at location $(u,v)$. In the global-QP setting, all locations share the same value; in the spatial-QP setting, the values vary across macroblock locations. We compute a soft assignment over a learnable set of quantization scaling factors $\{s_i\}_{i=1}^{L}$:
\begin{equation}
w_{u,v,i} =
\frac{
\exp\!\left(
-\frac{(q_{u,v}-i)^2}{\tau}
\right)
}
{
\sum_{j=1}^{L}
\exp\!\left(
-\frac{(q_{u,v}-j)^2}{\tau}
\right)
},
\label{eq:soft_weights}
\end{equation}

where $\tau$ controls the sharpness of the assignment. A smaller $\tau$ produces weights closer to hard index selection, while a larger $\tau$ yields smoother interpolation.

The resulting quantization scaling factor is obtained as

\begin{equation}
s_{u,v} = \sum_{i=1}^{L} w_{u,v,i} s_i.
\label{eq:soft_step_interp}
\end{equation}
For the global-QP setting, the formulation reduces to a single global scaling factor shared across the image. For the spatial-QP setting, the soft assignment is performed independently at each macroblock location, producing spatially varying scaling maps for encoder and decoder modulation. 
\begin{figure}[!h]
    \centering
    \includegraphics[width=1\linewidth]{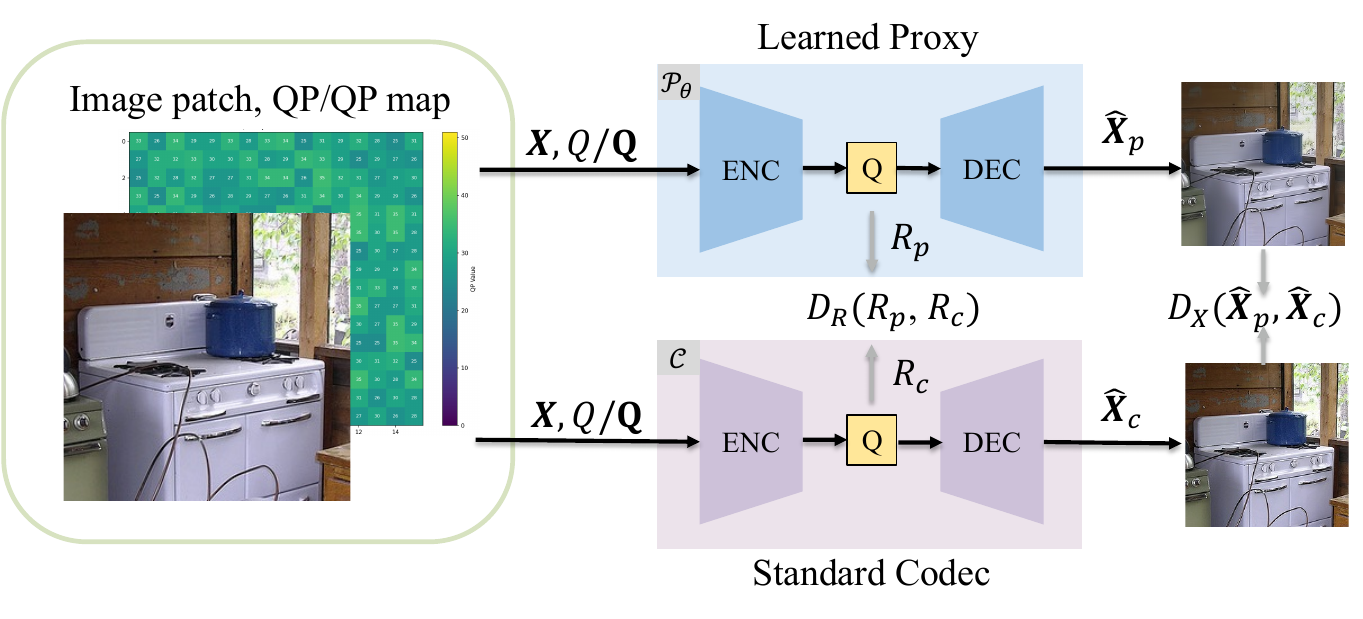}
    \caption{Training pipeline of the proposed differentiable proxy learning for H.264 Intra Codec}
    \label{fig:proxy}
\end{figure}

\begin{figure*}[!t]
    \centering
    \includegraphics[width=0.85\linewidth]{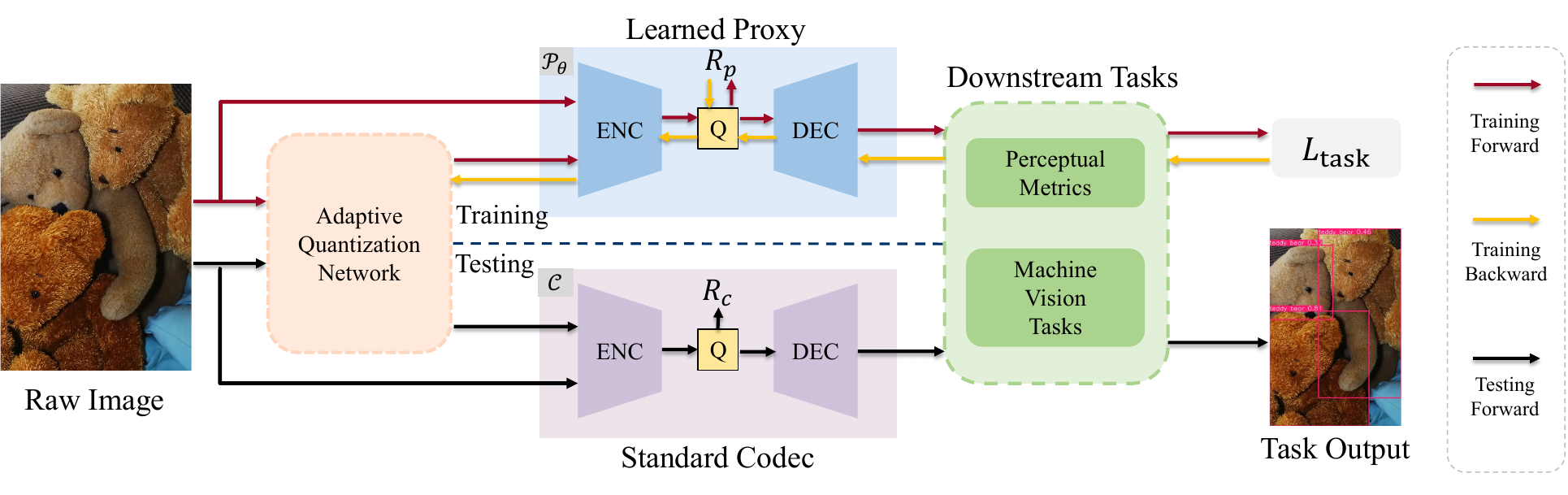}
    \caption{Proxy-based adaptive quantization control framework.}
    \label{fig:qpnet}
\end{figure*}

\subsubsection{Proxy training}



Fig.~\ref{fig:proxy} summarizes the proxy training pipeline. Given an input image $\mathbf{X}$, we first sample a QP $Q$ or QP map $\mathbf{Q}$ according to the target quantization setting. The same pair $(\mathbf{X},\mathbf{Q})$ is then passed to both the H.264 intra codec and the learned proxy. The codec produces the target reconstruction $\hat{\mathbf{X}}_c$ and bitrate $R_c$, while the proxy predicts $\hat{\mathbf{X}}_p$ and $R_p$. The proxy branch internally maps the codec-side QP configuration to its own control space before applying quantization scaling. The model is trained by matching both the reconstruction and bitrate of the standard codec.

For the global-QP proxy, the QP configuration reduces to a single scalar QP shared by the entire image. We predefine a set of $N$ anchor QP levels $\mathcal{Q}_{QP} = \{Q^{(1)}, Q^{(2)}, \ldots, Q^{(N)}\}$ to represent the target range of codec operating points. In our implementation, we use $\mathcal{Q}_{QP} = \{35, 40, 45, 51\}$. During training, we sample a QP from $\mathcal{Q}_{QP}$ and use the corresponding mapped proxy control parameter in the proxy branch.

For the spatial-QP proxy, we define a set of $M$ anchor QP distributions, $\mathcal{S} = \{ p^{(1)}, p^{(2)}, \ldots, p^{(M)} \}$ to cover a range of rate-distortion operating points. For each training image, we first sample one anchor QP distribution $p^{(m)} \in \mathcal{S}$, and then generate the entire QP map by independently sampling all macroblock QP values from that sampled anchor distribution: $Q_{u,v} \sim p^{(m)}, \quad \forall (u,v)$. 
In principle, the spatial QP map can follow an arbitrary distribution. In this work, we approximate this space using $M=5$ anchor uniform distributions, 
$\mathcal{S} = \{\mathcal{U}[20,30],\ \mathcal{U}[25,35],\ \mathcal{U}[30,40],\ \mathcal{U}[35,45],\ \mathcal{U}[40,51]\}$. The generated QP map $\mathbf{Q}$ is then applied to the H.264 codec to obtain the target reconstruction $\hat{\mathbf{X}}_c$ and bitrate $R_c$. Each element of $\mathbf{Q}$ is mapped to the corresponding proxy control parameter, producing a spatial quantization-control map $\mathbf{q}$.

The training objective for both global-QP and spatial-QP proxies is designed to match the rate-distortion behavior of the standard codec. We use a weighted combination of rate and distortion terms:

\begin{equation}
L_{p} = D_{R}(R_c, R_p) + \lambda \, D_{X}(\hat{\mathbf{X}}_c, \hat{\mathbf{X}}_p)
\label{eq:L1}
\end{equation}

where $D_R(\cdot)$ measures the bitrate prediction error (absolute difference in our implementation), and $D_X(\cdot)$ measures the reconstruction distortion (MSE in our implementation). The parameter $\lambda$ controls the trade-off between rate matching and distortion matching. Following DCVC-FM\cite{li2024neural}, $\lambda$ is calculated from $q$ as shown in Eq.~\ref{eq:lambda_interp}, and we further introduce an empirical multiplier $\alpha$ to better match some QP operating points.
\begin{equation}
\lambda = \alpha \cdot e^{\ln \lambda_{\min} + \frac{q}{q_{\text{num}} - 1} \cdot (\ln \lambda_{\max} - \ln \lambda_{\min})}
\label{eq:lambda_interp}
\end{equation}
where $[\lambda _{min}, \lambda _{max}]$ is set as $[1, 768]$, and $q_{\text{num}}$ is 64.

For the spatial-QP proxy, the training objective remains the same as in the global-QP case. However, since the quantization parameter varies across macroblocks, the Lagrange multiplier is computed from $\mathbf{q}$ and averaged over all macroblocks:
\begin{equation}
\bar{\lambda} = \frac{1}{H_b W_b} \sum_{u=1}^{H_b} \sum_{v=1}^{W_b} \lambda\!\left(q_{u,v}\right),
\end{equation}
where $q_{u,v}$ denotes the proxy control parameter at macroblock location $(u,v)$.

\subsection{Proxy-based adaptive quantization control}

Given the trained differentiable proxy $\mathcal{P}_{\boldsymbol{\theta}}$, we learn AQ networks that optimize QP configurations for specific downstream objectives. The networks are trained using the frozen proxy to provide differentiable rate and reconstruction signals, while inference is performed with the actual H.264 codec. The training and testing procedures are illustrated in Fig.~\ref{fig:qpnet}.

Let $\boldsymbol{\varphi}$ denote the parameters of an AQ network that predicts the QP configuration $\mathbf{Q}$ from input image $\mathbf{X}$. We optimize the AQ network to minimize a rate-task objective:
\begin{equation}
\boldsymbol{\varphi}^* = \arg\min_{\boldsymbol{\varphi}} \mathbb{E}\left[ R_p + \beta \, L_{\text{task}}(\hat{\mathbf{X}}_p) \right],
\label{eq:control_obj}
\end{equation}
where $R_p$ and $\hat{\mathbf{X}}_p$ are produced by the frozen proxy, $L_{\text{task}}$ is the downstream task loss (e.g., object detection loss or perceptual distortion), and $\beta$ controls the rate-task trade-off. This objective applies to both global-QP and spatial-QP settings, which differ in the granularity of quantization control and the parameterization of the AQ network.

\subsubsection{Adaptive global-QP control}

For adaptive global-QP control, we use a lightweight CNN-based AQ network to predict a single image-level QP value, enabling image-level bit allocation across different samples. Images that are more sensitive to downstream objectives can therefore be assigned lower QPs, while less sensitive images are compressed more aggressively. 

Specifically, the network consists of three strided convolution layers with output channels 16, 32, and 64, respectively, each followed by a ReLU activation. A global average pooling layer is then applied to aggregate spatial features, followed by a fully connected layer that regresses one scalar QP value. The output is mapped to the valid QP range using a sigmoid activation.

\subsubsection{Adaptive spatial-QP control}

\begin{figure}[!htbp]
    \centering
    \includegraphics[width=0.8\linewidth]{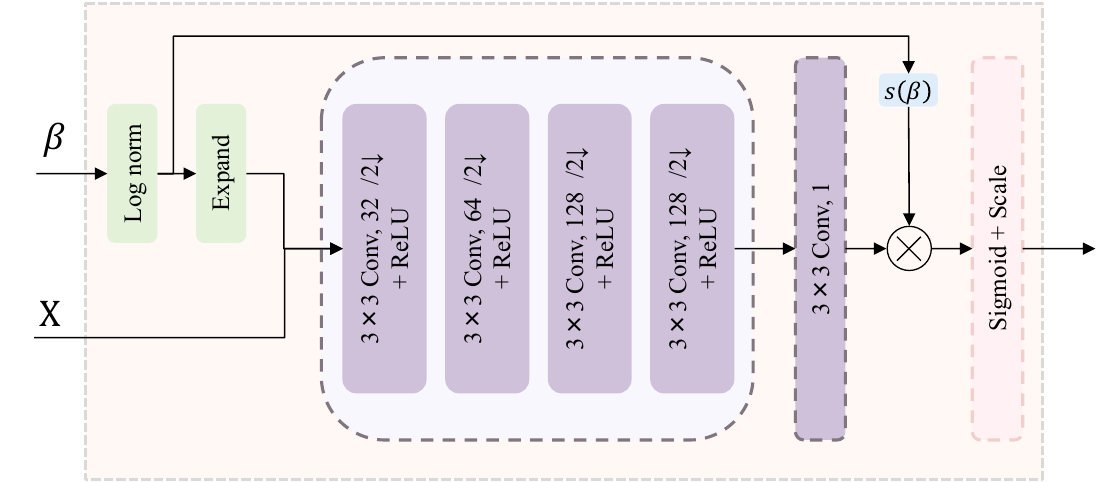}
    \caption{Structure of the AQ network for adaptive spatial-QP control.}
    \label{fig:aq-network}
\end{figure}

In the spatial-QP setting, the AQ network predicts a macroblock-level QP map for spatially varying bit allocation. As shown in Fig.~\ref{fig:aq-network}, the log-normalized trade-off parameter $\beta$ is expanded to an $H\times W$ map and concatenated with the input image. The resulting four-channel input is processed by a compact fully convolutional encoder to produce QP logits at the macroblock scale.

To incorporate the rate-task trade-off, we modulate the QP logits with a $\beta$-dependent factor
\begin{equation}
s(\beta)=1+\alpha\tanh(w\beta+b),
\end{equation}
where $w$ and $b$ are learned parameters and $\alpha=0.5$. The modulated logits are passed through a sigmoid and scaled to the valid QP range. This provides finer-grained adaptive quantization than global-QP control.

\section{Experiments}
In this section, we present the experimental results of the proposed differentiable proxy and the proxy-based AQ framework. We first evaluate the fidelity of the learned proxy to H.264 intra codec. Then, we demonstrate the effectiveness of the proxy-based codec control framework in improving rate-task trade-offs for both perceptual metrics and machine vision tasks.

\subsection{Evaluation of the proposed proxy for H.264 intra codec}
\subsubsection{Experimental setup}
The proxy was trained on the Vimeo-90k dataset \cite{xue2019video}, using only the first frame of each video sequence. Training data were generated by compressing I-frames with JM 19.0\cite{jm19} in YUV420 format under the anchor QP levels and distributions defined in Section~II-A. The compressed frames were converted to YUV444 format as input to the proxy. The model was implemented in PyTorch and trained using the Adam optimizer with a learning rate of $1\times10^{-5}$ and batch size of 4 on an NVIDIA 2080 Ti GPU.

The proxy was evaluated on the Kodak dataset \cite{kodakdataset} and compared against the H.264 reference model (JM 19.0) using PSNR and MS-SSIM metrics.

\subsubsection{Quantitative results}

\begin{figure}[!htbp]
    \centering
    \begin{subfigure}[b]{0.75\linewidth}
        \centering
        \includegraphics[width=\linewidth]{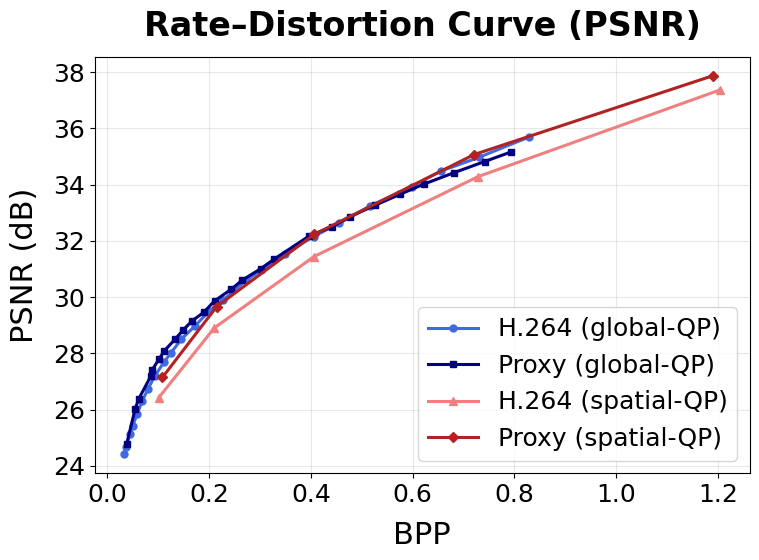}
        \label{fig:proxy-rd0-int}
    \end{subfigure}
    \begin{subfigure}[b]{0.75\linewidth}
        \centering
        \includegraphics[width=\linewidth]{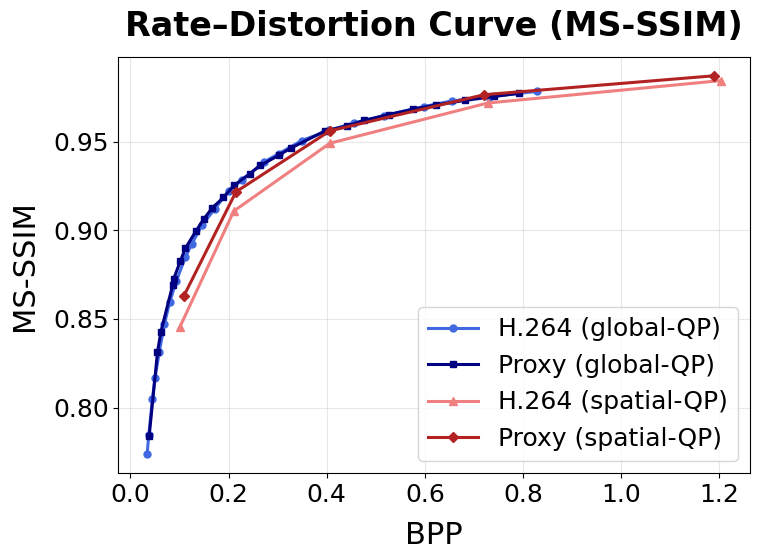}
        \label{fig:proxy-rd1-int}
    \end{subfigure}

    \caption{Rate-distortion curves of the global-QP and spatial-QP proxies in terms of PSNR and MS-SSIM tested on Kodak.}
    \label{fig:proxy-rd-all2}
\end{figure}

To evaluate the global-QP proxy, intermediate QP values within $[28, 51]$ were tested to obtain the corresponding rate-distortion curves. As shown in Fig. \ref{fig:proxy-rd-all2}, the selected anchor QP levels defined in Sec.~II-A cover the target range of rate-distortion operating points. Quantitatively, the global-QP proxy yields a BD-rate of $-4.75\%$ in terms of PSNR and $-2.56\%$ in terms of MS-SSIM.

\begin{figure*}[!tb]
    \centering
    \includegraphics[width=1\linewidth]{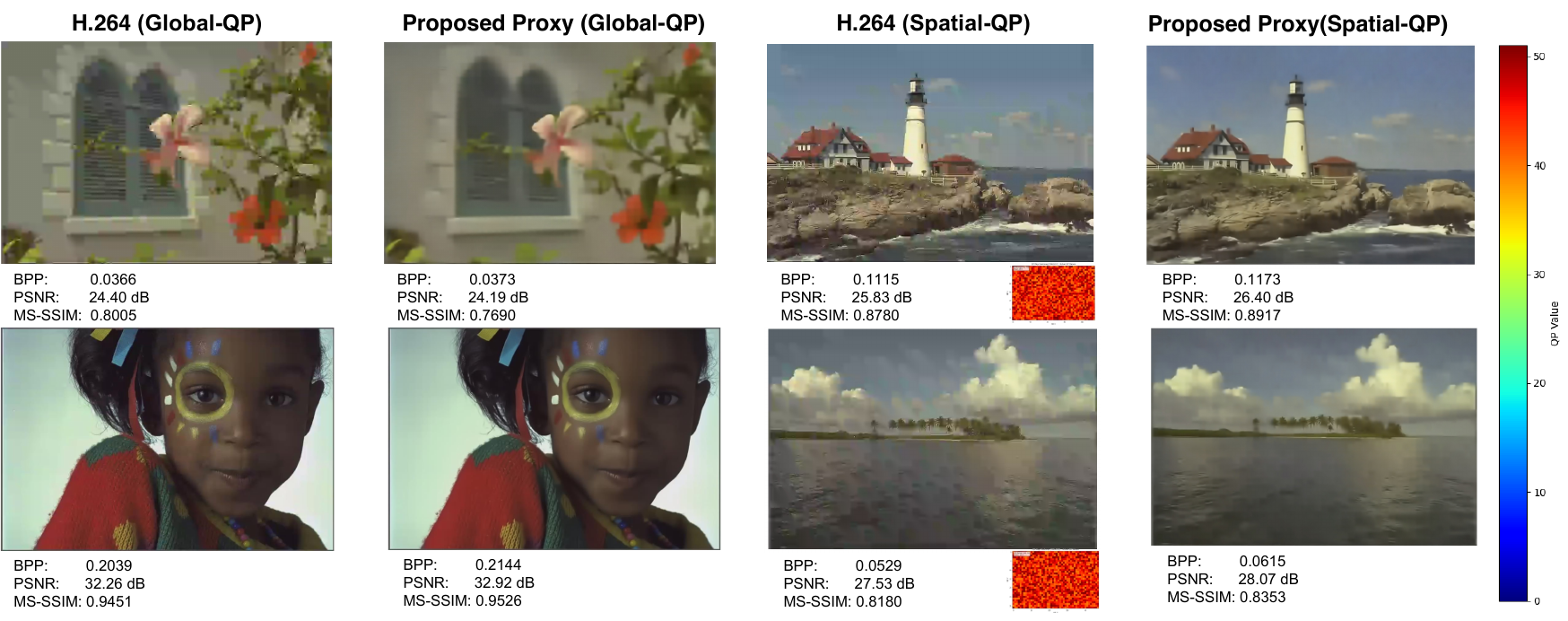}
    \caption{Visual comparison between H.264 and the proposed proxy under global-QP and spatial-QP settings.}
    \label{fig:res1}
\end{figure*}

To evaluate the spatial-QP proxy, we tested the model on the same set of anchor QP distributions defined in Sec.~II-A. As also shown in Fig. \ref{fig:proxy-rd-all2}, the proxy approximates the rate-distortion curves of the H.264 codec under different QP distributions, although the gap is larger than that of the global-QP proxy. Quantitatively, the spatial-QP proxy yields a BD-rate of $-15.01\%$ in terms of PSNR and $-12.91\%$ in terms of MS-SSIM. This result is expected because spatially varying quantization parameters introduce greater complexity and variability in the distortion patterns, making it more difficult for the proxy to capture all details accurately.

\subsubsection{Qualitative results}
As shown in Fig.~\ref{fig:res1}, the global-QP proxy produces reconstructions that are visually close to those of H.264, especially at relatively high bitrates. At lower bitrates, however, the proxy outputs remain noticeably smoother than the H.264 reconstructions. Compared with the strong blocking artifacts introduced by the standard codec, the proxy tends to generate more averaged textures.

For the spatial-QP proxy, the compressed results are generated using the random QP-map distributions illustrated in Fig.~\ref{fig:res1}. The proxy reproduces artifact patterns that roughly follow the corresponding spatial-QP allocation, indicating that it captures the effect of spatially varying quantization. Compared with H.264, the block-level transitions in the proxy outputs are less pronounced, and the reconstructed images remain relatively smoother overall.

\subsection{Evaluation of proxy-based adaptive quantization}

\subsubsection{Results on adaptive global-QP control}
For global-QP control, we train AQ networks on COCO to optimize machine vision and perceptual metrics, respectively. For machine vision experiments, we use YOLOv5\cite{jocher2022yolov5} with three model sizes (S/M/L). For each model size, $\beta$ is selected from $\{0.5, 1, 1.5, 2, 3, 4\}$ to obtain comparable bitrate operating points. The baseline is obtained with JM19.0 using a fixed QP across all images.
\begin{figure}[!htb]
    \centering
    \includegraphics[width=0.85\linewidth]{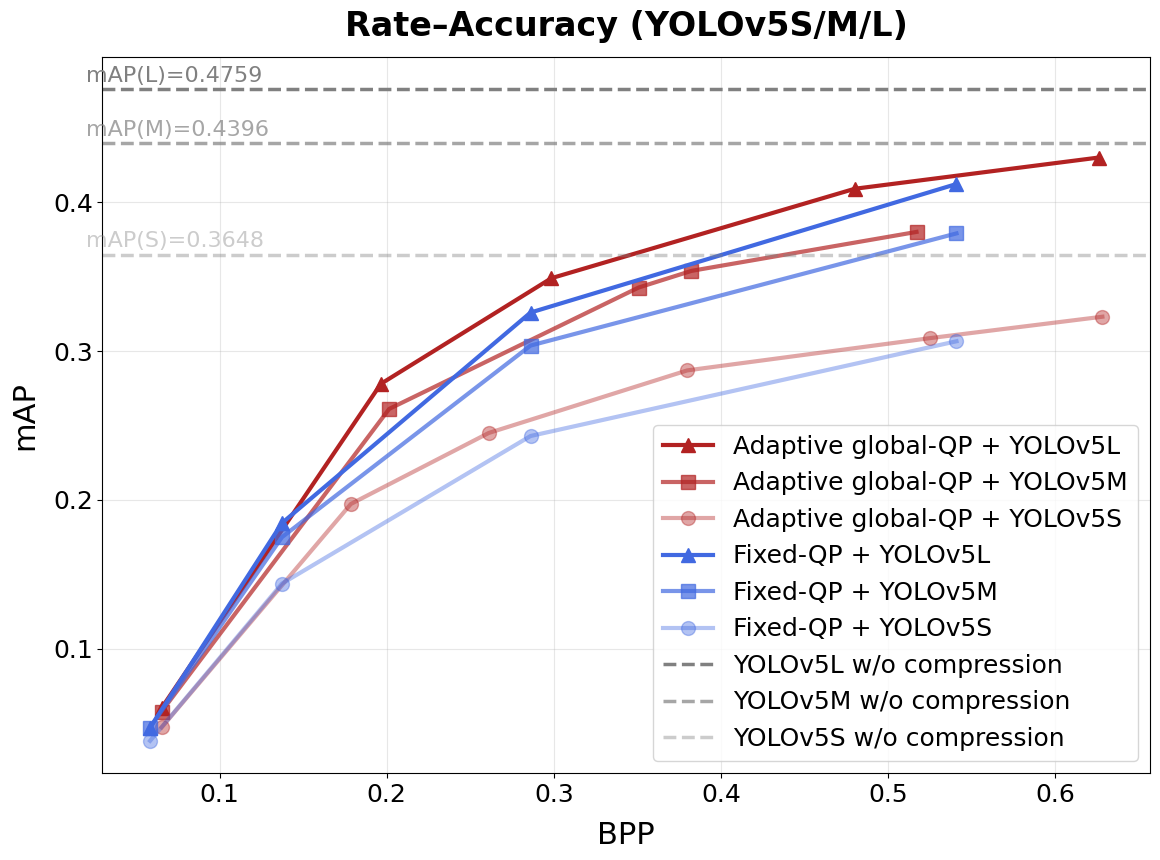}
    \caption{Rate-accuracy comparison on COCO using YOLOv5 backbones with three model sizes.}
    \label{fig:rate-mAP-single}
\end{figure}

As shown in Fig.~\ref{fig:rate-mAP-single}, the proposed framework consistently improves the rate-accuracy trade-off compared to the fixed-QP baseline across all three YOLOv5 model sizes on COCO. In terms of BD-rate, the adaptive global-QP control achieves savings of $-8.96\%$, $-8.19\%$, and $-9.50\%$ for YOLOv5S/M/L, respectively. 

\begin{figure}[h]
    \centering
    \includegraphics[width=0.75\linewidth]{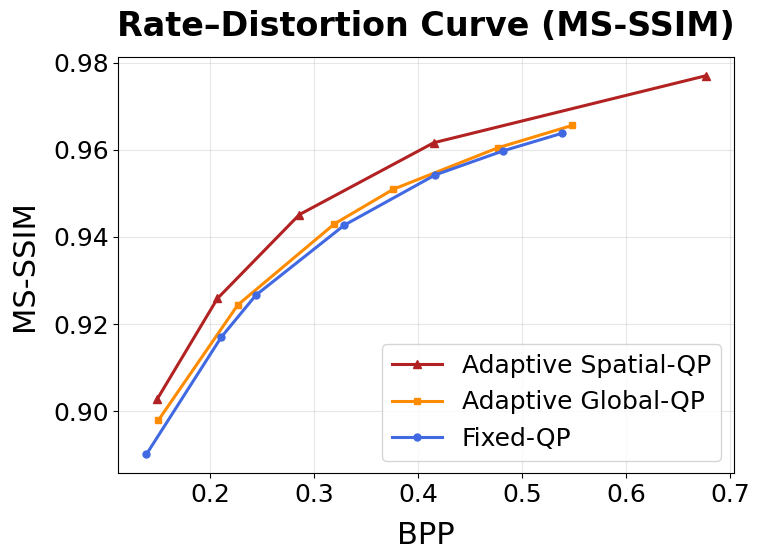}
    \caption{Rate--MS-SSIM comparison on Kodak.}
    \label{fig:rs-single}
\end{figure}

For the perceptual metric, the model is evaluated on Kodak in terms of rate--MS-SSIM. As shown in Fig.~\ref{fig:rs-single}, the proposed adaptive global-QP achieves a slightly better rate--distortion trade-off than the fixed-QP H.264 baseline, with a BD-rate of $-3.75\%$ in terms of MS-SSIM. This gain is relatively limited, suggesting that the room for improvement under image-level global-QP reallocation is smaller for perceptual optimization than for task-driven objectives such as object detection.
\subsubsection{Results on adaptive spatial-QP control}
\begin{figure*}[!t]
    \centering
    \includegraphics[width=0.75\linewidth]{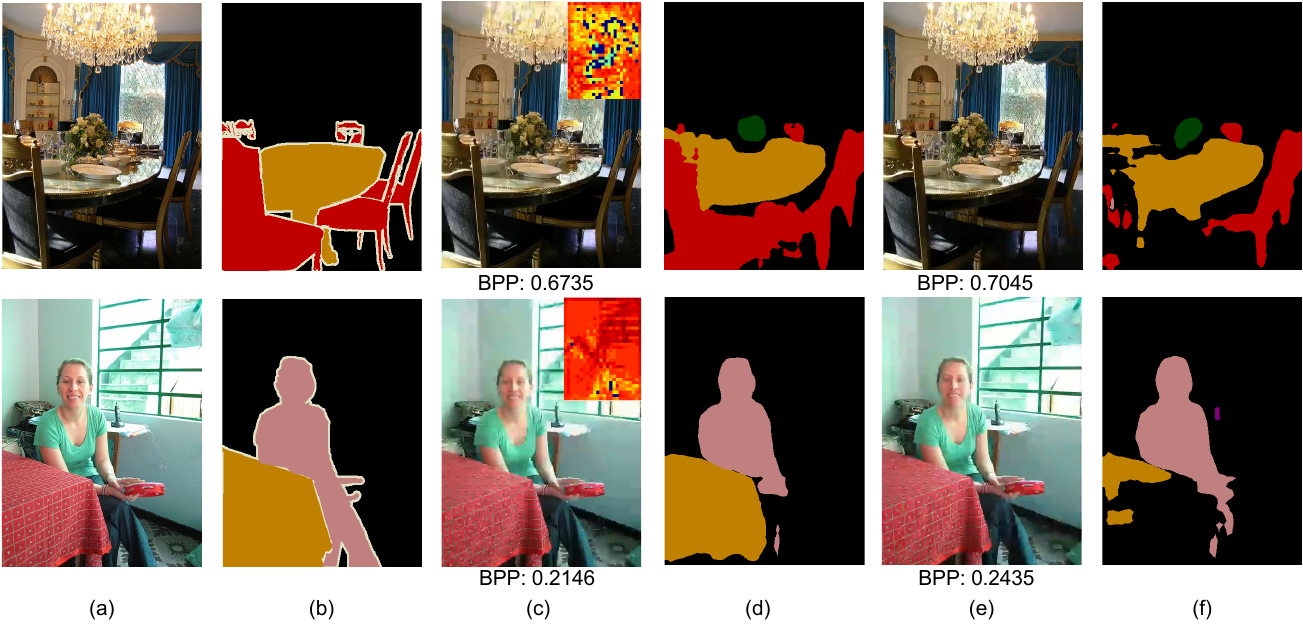}
    \caption{Visual comparison of segmentation results. (a) Original image. (b) Ground truth. (c) Reconstructed image using adaptive spatial-QP, with the QP map shown in the top-right corner. (d) Corresponding segmentation result. (e) Reconstructed image using fixed QP. (f) Corresponding segmentation result.}
    \label{fig:seg-visual}
\end{figure*}

\begin{figure}[!htbp]
    \centering
    \includegraphics[width=0.75\linewidth]{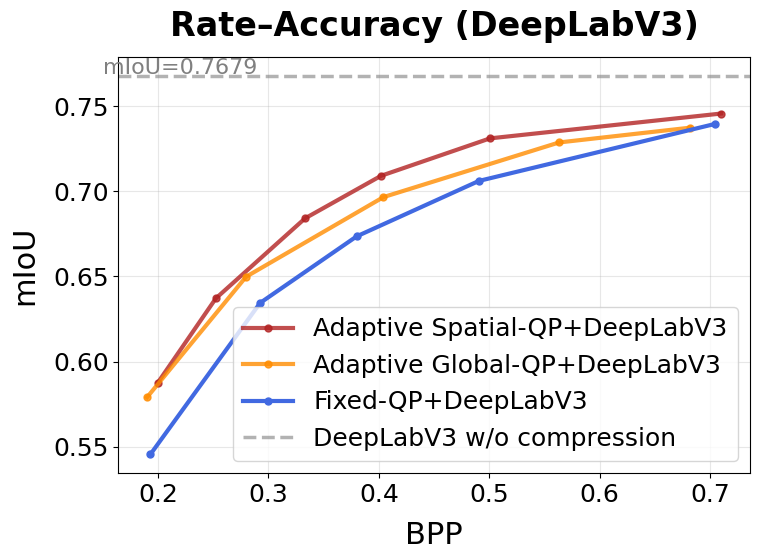}
    \caption{Rate-Accuracy comparison on Pascal VOC using Deeplabv3 with ResNet-50 backbone.}
    \label{fig:rd-multi-control}
\end{figure}

\begin{figure}[!htbp]
    \centering
    \includegraphics[width=0.7\linewidth]{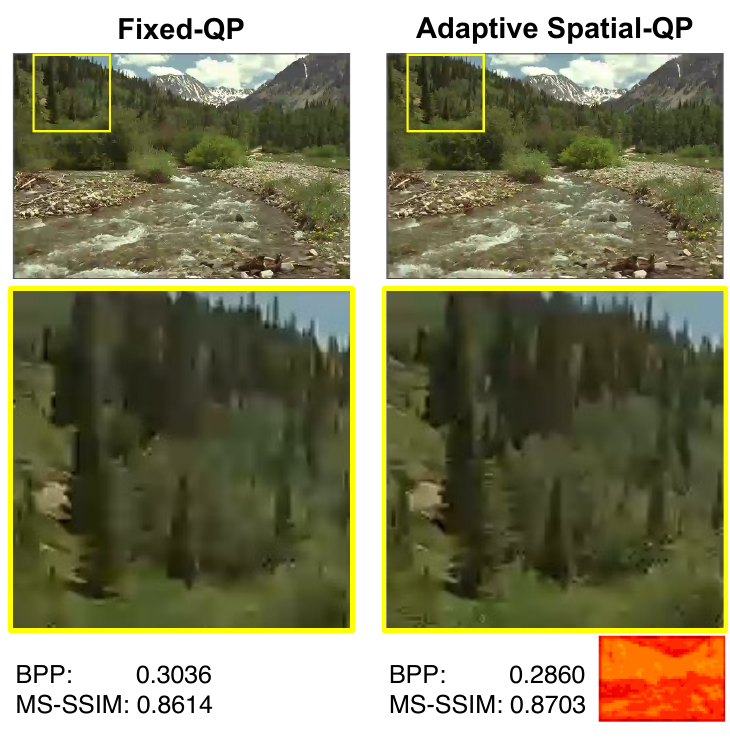}
    \caption{Visual examples of adaptive spatial-QP control optimized for MS-SSIM.}
    \label{fig:adaptive-spatial-vis}
\end{figure}

For adaptive spatial-QP control, we train AQ networks on Pascal VOC for semantic segmentation using DeepLabv3~\cite{chen2017rethinking} with a ResNet-50 backbone, and on Vimeo-90k for perceptual optimization using MS-SSIM as the objective. The corresponding rate--task results are summarized in Fig.~\ref{fig:rs-single} and Fig.~\ref{fig:rd-multi-control}. Qualitative examples are summarized in Fig.~\ref{fig:seg-visual} and Fig.~\ref{fig:adaptive-spatial-vis}.

For machine vision evaluation, we test on Pascal VOC and use JM19.0 with a fixed-QP as the baseline. As shown in Fig.~\ref{fig:rd-multi-control}, adaptive spatial-QP control achieves a BD-rate of $-17.12\%$, while adaptive global-QP control under the same segmentation setting achieves $-11.99\%$. This comparison shows that spatially varying QP prediction provides additional gains over image-level QP control for semantic segmentation. Visual examples are shown in Fig.~\ref{fig:seg-visual}, where adaptive spatial-QP preserves task-relevant regions with a lower bitrate than the fixed-QP baseline.

For the perceptual objective, adaptive spatial-QP control also improves rate--MS-SSIM performance, with a BD-rate of $-15.30\%$ (Fig.~\ref{fig:rs-single}). Qualitative examples are shown in Fig.~\ref{fig:adaptive-spatial-vis}. From a perceptual perspective, the predicted QP maps allocate more bits to visually salient structures and fewer bits to less sensitive regions, yielding cleaner textures and sharper structural details at comparable or lower bitrates.

\section{Conclusion}

In this paper, we presented a differentiable proxy learning method for H.264 intra codec to enable adaptive quantization for different downstream tasks. Built upon a variable-rate learned compression model, the proposed proxy supports both global-QP and spatial-QP settings through a differentiable soft-indexing mechanism. Experimental results showed that the learned proxy can approximate the rate-distortion behavior of H.264 intra codec under different quantization settings.

Using the frozen differentiable proxy, we further developed proxy-based AQ frameworks for perceptual optimization and machine vision tasks. The proposed framework consistently improved rate-task trade-offs over fixed-QP H.264 baselines in object detection, semantic segmentation, and perceptual optimization experiments. In particular, spatial-QP control demonstrated additional gains over image-level global-QP control by enabling finer-grained bit allocation across image regions. Our results also suggest that highly accurate proxy reconstruction is not strictly necessary for effective AQ optimization, as meaningful rate-task improvements can still be achieved under imperfect proxy fidelity. Extending the proposed framework to inter-frame video coding with temporal prediction remains an important direction for future work.

\bibliographystyle{IEEEbib}
\bibliography{references}

\end{document}